\documentclass[structabstract]{aa}
\usepackage{txfonts}
\usepackage{graphicx}
\usepackage{natbib}
\usepackage[dvips]{color}
\usepackage{widetext}

\bibpunct{(}{)}{;}{a}{}{,}

\begin{document}

\newcommand{\be}{\begin{equation}}
\newcommand{\ee}{\end{equation}}
\newcommand{\bdm}{\begin{displaymath}}
\newcommand{\edm}{\end{displaymath}}
\newcommand{\bea}{\begin{eqnarray}}
\newcommand{\eea}{\end{eqnarray}}

\newcommand{\cf}{\textit{cf.}~}
\newcommand{\ie}{\textit{i.e.}~}

\newcommand{\vm}[1]{ \textcolor{blue}  {\texttt{\textbf{VM: #1}}} }
\newcommand{\oz}[1]{ \textcolor{red}   {\texttt{\textbf{OZ: #1}}} }

\title{Particle acceleration in the polar cap region of
  an oscillating neutron star}

\author{
        O. Zanotti    \inst{1,2}
\and    V. Morozova   \inst{2,3,4,5}
\and    B. Ahmedov \inst{3,4}
}

\institute{
Laboratory of Applied Mathematics, University of Trento,
Via Mesiano 77, I-38100 Trento, Italy
\\ \email{olindo.zanotti@gmail.com}
\and
Max-Planck-Institut f{\"u}r Gravitationsphysik, Albert Einstein Institut, Golm, Germany
\and
Institute of Nuclear Physics,
        Ulughbek, Tashkent 100214, Uzbekistan
\and
Ulugh Beg Astronomical Institute,
        Astronomicheskaya 33, Tashkent 100052, Uzbekistan
\and
ZARM, University Bremen, Am Fallturm, 28359 Bremen, Germany
}

\date{}

\authorrunning{O.~Zanotti et al.}
\titlerunning{Particle acceleration in NS magnetospheres}

\abstract
{}
{We revisit particle acceleration in the
  polar cap region of a neutron star by taking into
  account both general relativistic effects and the
  presence of toroidal oscillations at the star surface.
In particular, we
address the question of whether toroidal oscillations at the stellar
surface can affect the acceleration properties in the polar cap.
}
{We solve numerically the
relativistic electrodynamics equations in the stationary regime,
focusing on the computation of the Lorentz factor of a
space-charge-limited electron flow
accelerated in the polar cap region of a
rotating and oscillating pulsar. To this extent,
we adopt the correct expression of the general
relativistic Goldreich-Julian charge 
density in the presence of toroidal oscillations.
}
{
Depending
on the ratio of the actual charge density of the pulsar
magnetosphere to the Goldreich-Julian charge density,
we distinguish
two different regimes of  the Lorentz factor of the
particle flow,
namely an oscillatory regime produced for
sub-GJ current density configurations,
which does not produce an
efficient acceleration, and a true accelerating regime
for super-GJ current density configurations.
We find
that star oscillations
may be responsible for a significant asymmetry in the pulse
profile that depends on the orientation of the oscillations with
respect to the pulsar magnetic field. In particular,
significant enhancements of the Lorentz factor are
produced by stellar oscillations in the super-GJ current density regime.
}
{}


\maketitle


\section{INTRODUCTION}
\label{Introduction}

Understanding the physical mechanism behind particle acceleration, and
hence electromagnetic emission, in the magnetospheres of
neutron stars remains an unsolved topic of pulsar physics,
more than forty years after their discovery. A completely
force-free (FF) magnetosphere with no parallel component of the
electric field along the magnetic field lines, clearly does not allow for
the acceleration of particles, and local violations to the FF
condition are therefore expected~\citep{Kalapotharakos2011}.
This was first noticed by \citet{Sturrock1971},
who
proposed that the acceleration of particles
occurs in a small vacuum gap, where $\vec
E\cdot\vec B\neq0$, that is confined to the polar region above
the star surface, the so-called polar cap, whose edge is
defined by the locus of the last closed magnetic field
line.
This seminal idea was developed and 
investigated by several authors over the years,
both in flat spacetime~\citep{Ruderman1975,
Fawley1977,Scharlemann1978,Arons1979,Cheng1980,Daugherty1982,Shibata1997} and in a
general relativistic
framework~\citep{Muslimov1992,Muslimov1997,Harding1998,Sakai2003,Morozova2008}.

A relevant result obtained by these works is that the
particle motion can show an oscillatory behavior
depending on the charge current density.
This behavior has been indirectly confirmed by
numerical simulations. 
By performing time-dependent particle-in-cell simulations of
electrons extracted from the star surface, \cite{Barzilay2011} 
found that the Lorentz factor of the accelerated electrons
reaches the analytical estimates of $10^6$ and follows an oscillatory
pattern that increases with the distance from the stellar
surface.
Although these numerical analyses are very promising, 
numerical simulations of neutron star
magnetospheres have not  yet entered  a mature enough
stage where physical mechanisms such as plasma effects,
non-ideal magnetohydrodynamics, pair-creation, and
radiation processes can be properly taken into account.
As a result, analytical investigations continue to play
a crucial role in elucidating the physics of pulsar
magnetospheres.

Among the physical effects deserving particular attention is
the possibility that neutron
star oscillations, most likely excited during a glitch phenomenon
(i.e. a sudden change in the rotational period), propagate into the
magnetosphere, thus affecting the acceleration properties in the
polar cap region.
The idea that neutron star oscillations may induce high
energy emission in neutron star magnetospheres was developed
in a series of papers by Timokhin and collaborators
\citep{Timokhin2000,Timokhin2007,Timokhin2008}, after the
first pioneering investigations by \cite{McDermott1984},
\cite{Muslimov1986}, and \cite{Rezzolla2004},
who considered the case of an oscillating neutron star in a
vacuum.
A few years ago we initiated a program aimed at assessing the impact of
toroidal oscillations of the star surface on the
properties of the
external magnetosphere, with general relativistic effects
properly taken into account.
Starting from \cite{Abdikamalov:2008sk},
where the theoretical basis of our approach was
developed, we showed in
\cite{Morozova2010}  that the electromagnetic energy losses
from the polar cap region of a rotating neutron star can be
significantly enhanced if oscillations are also present, and, for
the mode $(l,m)=(2,1)$, these electromagnetic losses turned out to be
a factor $\sim8$ larger than the rotational energy losses, even for
a velocity oscillation amplitude at the star surface as small as
$\tilde\eta=0.05 \ \Omega \ R$, where $\Omega$ and $R$ are
  the angular velocity and the radius of the
neutron star, respectively.
In \cite{Morozova2012}, on the
other hand, we  considered the conditions
for radio emission in magnetars
and we found that, when oscillations of the
magnetar are taken into account, radio emission from the
magnetosphere is generally favored. The major effect of
oscillations is to amplify the scalar potential in the polar cap
region of the magnetar magnetosphere, which implies that the
death-line in the $P-\dot{P}$ diagram shifts
downward\footnote{However, we also found that
when the compactness of the neutron star is
increased, the death line shifts
upwards in the $P-\dot{P}$ diagram, pushing the
magnetar into the radio-quiet region and explaining
the observational evidence that most of the
known magnetars are radio-quiet.}.

As a natural follow-up of our investigations,  we
address in this paper the question of whether toroidal oscillations at the star
surface can affect the acceleration properties in the polar cap
region. We rely on our results presented in
\cite{Morozova2010} revisiting some of the
conclusions found by \cite{Sakai2003}.  In
  particular, we expect that an oscillating magnetosphere
can alter the parallel component of the electric field
that is responsible for particle acceleration just above
the star surface. The size of oscillation amplitude
necessary to produce significant physical effects is still a
matter of debate. \cite{Timokhin2008}, for example, argue that
noticeable effects in the spectra of
soft gamma repeater giant flares can be produced
when the amplitude of the neutron star surface
oscillations is  of the order of $\sim 1\%$ of the star
radius. As we show in the paper, on the other hand,
significant effects on the Lorentz factor are visible
even with (normalized) velocity oscillation amplitudes
$K=\tilde\eta/\Omega R$ as low as $0.02$.
Since the range of oscillation
frequencies is quite broad, typically between $18$ Hz and
$1800$ Hz~\citep{Steiner2009}, such low values of $K$ can
be obtained, for example, by adopting an oscillation
frequency that is $20$ times higher than the stellar
rotation frequency $\Omega$, and a spatial oscillation
amplitude that is only $0.001R$.
We anticipate, therefore, that the effects of an
oscillating magnetosphere on the acceleration of
particles close to the stellar surface become important, even
for very small oscillation amplitudes.

The plan of the paper is the following. In Sec.~\ref{The_Physical_model},
we describe the basic features of the physical model, while
Sec.~\ref{Results} is devoted to the presentation and 
discussion of the results. Sec.~\ref{Conclusions}
contains the conclusions of our work.
 In addition to the Gauss units for the magnetic field, we set
$c=G=1$. In this way, any quantity, including the magnetic field, is
geometrized, namely there is only one unit of measure, which is the
gravitational radius $r_g=GM/c^2$.
We keep $c$ and $G$ in an explicit form in
those expressions of particular physical
interest. Appendix~\ref{appendixA} describes the extended geometrized
system of units adopted in the numerical solution of
Eqs.~(\ref{system1}) and (\ref{system2}).

\section{THE PHYSICAL MODEL}
\label{The_Physical_model}

\subsection{Particle dynamics in the pulsar polar cap}
\label{Particle_dynamics_in_the_pulsar_polar_cap}

We assume that the space-time outside a representative
neutron star of mass $M$, radius $R$, and angular
velocity $\Omega$
is given by the Hartle \& Thorne metric in
the slow-rotation limit~\citep{Hartle1968}, i.e.
\bea ds^2&=&-N^2dt^2+N^{-2}dr^2-2\omega_{\rm LT} r^2\sin^2\theta d\phi dt
+ \nonumber \\
&& r^2(d\theta^2+\sin^2\theta d\phi^2)\ ,
\eea
where $N=\sqrt{1-\frac{2GM}{c^2r}}$ is the lapse function,
$\omega_{\rm LT}=\frac{2\tilde{J}}{r^3}$ is the Lense-Thirring angular
velocity, and $\tilde{J}=I\Omega$ is the total angular momentum
of the neutron star measured from infinity, with $I$
being the moment of inertia.
The frozen dipole-like magnetic field of the pulsar has
the following orthonormal components in spherical coordinates
$(r,\theta,\phi)$~\citep[see, for instance,][]{Ginzburg1965,Wasserman1983, Rezzolla2001}
\bea
\label{Br}
B_{\hat{r}}&=&B_0
\frac{f(\eta)}{f(1)}\frac{1}{\eta^3}\cos\theta , \\
\label{Btheta}
B_{\hat{\theta}}&=&\frac{1}{2}B_0
N\left[-2\frac{f(\eta)}{f(1)}+\frac{3}{\left(1-\frac{\varepsilon}{\eta}\right)f(1)}\right]
\frac{1}{\eta^3} \sin\theta ,
\eea
where
\be \label{f}
f(\eta)=-3\left(\frac{\eta}{\varepsilon}\right)^3\left[\ln\left(1-\frac{\varepsilon}{\eta}\right)+\frac{\varepsilon}{\eta}
\left(1+\frac{\varepsilon}{2\eta}\right)\right]\ .
\ee
In the expressions above,
$\eta=r/R$ is the normalized radius, $\varepsilon=2M/R$ is
the compactness parameter of the neutron star, and $B_0$ is the
value of the magnetic field at the pole of the star, which
is of the order of $10^{12}\mathrm{G}$
for a typical pulsar.
Over the years, a series of fundamental works
\citep[see, among 
  others,][]{Goldreich1969,Ruderman1975,Arons1979,Mestel1985,Fitzpatrick1988,
Michel1991,
  Muslimov1997,Beskin2010}
have helped to reveal the
mechanism behind the pulsar magnetosphere generation.
A strong surface magnetic field and
a relatively rapidly rotating pulsar lead to the
generation of an  electric field in the vicinity of the
pulsar.
In the framework of a space-charge-limited-flow
  accelerator that we  assume here,
the parallel component of the electric field $E_\|$ at the
stellar surface is zero~\citep{Becker2009}.
Therefore, if the surface temperature
is higher than the thermionic emission temperature,
particles can leave the star, but
with non-relativistic speeds. Once
above the star surface, electrons are accelerated
to ultra-relativistic velocities on a short distance scale
by the non-zero $E_\|$.
This first generation of electrons, called
primary electrons,
emit curvature $\gamma$-quanta,
which, in turn, create electron-positron pairs, initiating the
cascade process that forms the pulsar magnetosphere.

The charge density of the pulsar magnetosphere, which is required in
order to completely screen the electric field component parallel to
the magnetic field of the pulsar, is called the Goldreich-Julian charge
density.
The actual charge density $\rho$ of the pulsar magnetosphere
slightly deviates from the Goldreich-Julian charge
density and the deviation can be roughly estimated from
the one-dimensional equation~\citep{Beskin2010}
\be
\frac{dE_\|}{dh}=\rho-\rho_{\rm GJ, rot}\,,
\ee
where $h$ is the distance from the star surface.
The expression for the Goldreich-Julian
charge density of the pulsar, corrected for general
relativistic effects, was obtained by
\citet{Muslimov1992} and  is given by
\bea \label{inclined_rho} && \rho_{\rm GJ, rot}= \nonumber \\
&& \quad -\frac{\Omega B_0}{2\pi
c}\frac{1}{N\eta^3}\frac{f(\eta)}{f(1)}\left[\left(1-\frac{\kappa}{\eta^3}\right)\cos\chi
+\frac{3}{2}H(\eta)\theta\sin\chi\cos\phi\right]\ , \eea
where
\be \label{H}
H(\eta)=\frac{1}{\eta}\left(\varepsilon-\frac{\kappa}{\eta^2}\right)+
\left(1-\frac{3\varepsilon}{2\eta}+\frac{\kappa}{2\eta^3}\right)\left[f(\eta)\left(1-\frac{\varepsilon}{\eta}\right)\right]^{-1}\
,
\ee
while $\chi$ is the inclination angle between the axis of rotation
of the pulsar and the magnetic moment of the dipole.
Moreover, $\kappa\equiv\varepsilon\beta=(\omega_{\rm LT})_{\ast}/\Omega$,
where the asterisk denotes a value at the surface of the
star,
and $\beta=I/I_0$ is the normalized moment of inertia,
where $I_0=MR^2$. Since the polar cap
  is located in the vicinity of the magnetic pole,
in Eq.~(\ref{inclined_rho}) the polar axis of the spherical
coordinates $(r,\theta,\phi)$ is chosen to be directed along the
magnetic dipole axis, which also explains why the angle
$\chi$ does not appear in the expressions for the
magnetic field components given in Eqs.~(\ref{Br}) and (\ref{Btheta}).

Within this framework, it is possible to solve the
equations of motion of the electron
accelerated in the vicinity of the polar cap.
Under the
assumption that the charge flow reaches a stationary
configuration,
\citet{Shibata1997} and \citet{Sakai2003} formulated this
problem in the form
\bea \label{system1}
&&\frac{N}{s^2}\frac{d}{ds}\left(s^2\frac{d\phi}{ds}\right)-\frac{l^\prime(l^\prime+1)}{Ns^2}\phi=\frac{B}{B_0N}\left(\frac{j}{v}-\bar{j}\right)
\ ,
\\ \label{system2}
&&\frac{d}{ds}(N\gamma)=\frac{1}{N^2}\frac{d\phi}{ds} \ ,
\eea
where
\bea
\label{radial_coord}
s\equiv\sqrt{\frac{2\Omega B_0 e}{m c^3}}r
\eea
is a normalized radial coordinate, while $v$ is the
velocity of the electron. In equations (\ref{system1})
and (\ref{system2}),
additional normalizations are performed as
\bea
\label{variables}
j&\equiv&-\frac{2\pi N(s) J(s)}{\Omega B(s)} \ , \\
\phi(s)&\equiv&\frac{e}{m}\Phi(s)\ , \\
\bar{j}&\equiv&-\frac{2\pi N(s)~ c ~\rho_{\rm GJ,rot}(s)}{\Omega B(s)}\approx
\nonumber \\
&&\left(1-\frac{\kappa}{\eta^3}\right)\cos\chi+\frac{3}{2}H(\eta)\theta\sin\chi\cos\phi
\ ,
\eea
where $\mathbf{J}$ is the current density in the pulsar
magnetosphere, $\Phi$ is the scalar electric potential, $m$ and $e$ are the
mass and the charge of the electron, respectively, $\gamma$ is the
Lorentz factor of the accelerated electrons and
$B\equiv|\mathbf{B}|=(B^2_{\hat{r}}+B^2_{\hat{\theta}})^{1/2}\approx
B_{\hat{r}}+{\cal O}(\theta^2)$.
Moreover, we stress
that, in order to derive the equation (\ref{system1}) from the relativistic
Poisson equation, the scalar potential $\Phi$ has been expanded in the
angular directions as
\be
\label{scalar_potential}
\Phi=\sum_{l^\prime,m^\prime}\bar{\Phi}Y_{l^\prime m^\prime}(\theta,\phi)\,,
\ee
where the spherical orthonormal functions
$Y_{lm}(\theta,\phi)$ are the eigenfunctions of the Laplacian
operator in spherical coordinates.
Only modes of the polar cap scale have been considered, with
$l^\prime\approx\pi/\Theta_0$, where
\be  \Theta_0=\arcsin\left[\frac{R}{R_{LC}f(1)}\right]^{1/2}  \ee
is the polar angle of the last open magnetic field line at the
surface of the star~\citep{Muslimov1992},
while $R_{LC}=c/\Omega$ is the radius of the light cylinder.
In our computations,
for typical values of
$\Theta_0$ in the range $\approx[2^\circ,3^\circ]$,  we
have $l^\prime\approx[60, 90]$.
The polar trajectory of the
accelerated particles is described by the curve
\be \theta(s)\approx\theta_\ast\sqrt{\frac{s}{s_\ast}}\ , \ee
and remains
bounded by the polar angle of the last open
magnetic field line~\citep{Morozova2010}, i.e.
\be
\theta(s)\leq
\Theta\cong\arcsin\left\{\left[\eta\frac{f(1)}{f(\eta)}\right]^{1/2}\sin\Theta_0\right\}\ .
 \ee

\subsection{The effects of pulsar oscillations}

In our analysis, we also
consider the case when the rotating pulsar is subject to
toroidal oscillations, with the orthonormal
velocity field given by~\citep{Unno1989}
\be \label{vel} \delta v^{\hat{i}}=\left\{0,
\frac{1}{\sin\theta}\partial_{\phi}Y_{l m}(\theta,\phi),
-\partial_{\theta}Y_{lm}(\theta,\phi)\right\}\tilde{\eta}(r)e^{-i\omega
t}\ , \ee
where $\omega$ is the real part of the oscillation
frequency and 
$\tilde{\eta}$ is the radial eigenfunction expressing the amplitude
of the oscillations. We assume that the oscillations
of the stellar crust can induce oscillations in the plasma
magnetosphere, at least in the close vicinity of the surface, which
is our main region of interest. As a result, in the rest of our
analysis we neglect the radial dependence of the oscillation
amplitude, i.e. we assume $\tilde{\eta}(r)=\tilde{\eta}(R)$.

In principle, an additional degree of freedom should be
introduced, related to the possible uncertainty in 
the orientation of the
oscillations with respect to both the axis of rotation and
the pulsar magnetic field. However,
we adopted here a  simplified approach and, as
implied by Eq.~(\ref{vel}), 
oriented the oscillation mode axis
along the $z-$ axis of the magnetic dipole,
while the rotational axis is inclined.

The Goldreich-Julian charge density of a rotating and
oscillating pulsar was obtained by~\citet{Morozova2010}
and is given by\footnote{We were careful 
  not to confuse $l^\prime$ and $m^\prime$ (which are used in
  the expansion of the scalar potential
  in Eq.~(\ref{scalar_potential})), with $l$ and $m$ used in the
  expansion of the perturbation in Eq.~(\ref{vel}).}
\bea \label{rhoGJfin} && \rho_{\rm{GJ}}=\rho_{\rm
  GJ,rot}+\rho_{\rm GJ,osc}= \nonumber \\ && \qquad
\rho_{\rm GJ,rot} - \frac{1}{4\pi
c}\frac{1}{R\eta^4}\frac{B_0 e^{-i\omega
t}}{\Theta^2(\eta)}\frac{1}{N}\frac{f(\eta)}{f(1)}\tilde{\eta}(1)l(l+1)Y_{l
m} \ . \eea
After using the expression above,
one can easily obtain
the current $\bar{j}$ in the presence of both rotation
and oscillations as
\bea \label{barj} &&
\bar{j}=\left(1-\frac{\kappa}{\eta^3}\right)\cos\chi+\frac{3}{2}H(\eta)\theta\sin\chi\cos\phi+
\nonumber \\ && \qquad\qquad
\frac{1}{2}\left(\frac{f(\eta)}{f(1)}\right)^{\frac{2-m}{2}}\frac{\theta^m}{\Theta_0^2}
K\eta^{\frac{m}{2}-2}l(l+1)A_{lm}\ , \eea
where the amplitude of the oscillation has been
 parametrized in terms of the small number
$K=\tilde{\eta}(R)/\Omega R$, giving the ratio of the velocity
of oscillations to the linear rotational velocity of
neutron star.
For any particular mode $(l,m)$, we used
the approximation $Y_{lm}(\theta,\phi)\approx
A_{lm}(\phi)\theta^m$, which is valid in the limit of small polar angles
$\theta$, where the terms $A_{lm}(\phi)$ appearing in Eq.~(\ref{barj})
have
real parts given by
\begin{eqnarray}
\label{listA}
A_{00}&=&\frac{1}{\sqrt{4\pi}}\ , \\
A_{10}&=&\sqrt{\frac{3}{4\pi}} \ , \\
A_{11}&=&-\sqrt{\frac{3}{8\pi}}\cos\phi \ , \\
A_{20}&=&\sqrt{\frac{5}{4\pi}} \ , \\
A_{21}&=&-3\sqrt{\frac{5}{24\pi}}\cos\phi\ .
\end{eqnarray}

We note that in all our analysis we only consider constant ratios
$j/\bar{j}_\ast$, namely we
assume that the ratio
of actual charge density to the Goldreich-Julian charge density is
constant.
This should be regarded as a very good
approximation as long as the acceleration mechanism is
confined to a short distance from the stellar surface.

We stress that the two most plausible models 
of pulsar magnetosphere, namely those of \cite{Mestel1981} and
\cite{Arons1979},
predict different distributions of the
charge density in the polar cap region, and hence different
conditions for pair creation. 
In the model of Mestel, a super-GJ charge density is formed
along the magnetic field lines that  curve away from the
rotational axis, while along the field lines curving towards the
rotational axis a perfect screening of the accelerating electric
field component can be  achieved and the charge density is equal to the
Goldreich-Julian charge density. 
In contrast, in the model of Arons,
the region where magnetic field lines
curve towards the rotational axis has a super-GJ charge
density near the star surface, and a sub-GJ
charge density away from it\footnote{On the
  other hand, this type of solution does not apply in the region where
magnetic field lines curve away from the rotational
axis.}.  
One of the main differences between the two
models, which is responsible for  different predictions about 
the charge density in the pulsar magnetosphere, lies in
the boundary conditions of the problem: in
the model of Mestel, in particular,
the charge density is taken to be equal to the Goldreich-Julian
charge density at the surface of the star, while Arons suggests
that $E_{\|}=0$ far away from the surface.

However, the present lack of a fully self-consistent model of 
pulsar magnetospheres motivates a pragmatic approach,
according to which the charge density can be treated as a
free parameter.

\section{RESULTS}
\label{Results}

In a first series of calculations, we have solved the system
of equations given by
Eqs.~(\ref{system1})-(\ref{system2}) for the case of a
rotating but 
non-oscillating pulsar. The two equations Eqs.~(\ref{system1}) and
(\ref{system2}) have been reduced to a system of three ODEs in the
unknowns $\phi$, $d\phi/ds$, and $N\gamma$, which has been solved
with a standard fourth-order Runge-Kutta method. The main parameters
of our fiducial pulsar model are $R=10 \
\mathrm{km}$, $M=1.4 M_\odot$, $\Omega=2\pi/(0.1 \mathrm{s})$,
$B_0=1.0\times10^{12} \mathrm{G}$, $\kappa=0.1$, and $\chi=30^\circ$.
We note that, by selecting $\kappa=0.1$, the moment of inertia of the
pulsar is given by $I/I_0\approx0.24$. The initial value of the
Lorentz factor at the star surface is $\gamma_\ast=1.01$, while the
initial $\theta_\ast$ is bounded by $\Theta_0$, namely
is $0\leq\theta_\ast\leq\Theta_0$.

\begin{figure}
\centering
{\includegraphics[angle=0,width=8.3cm,height=7.5cm]{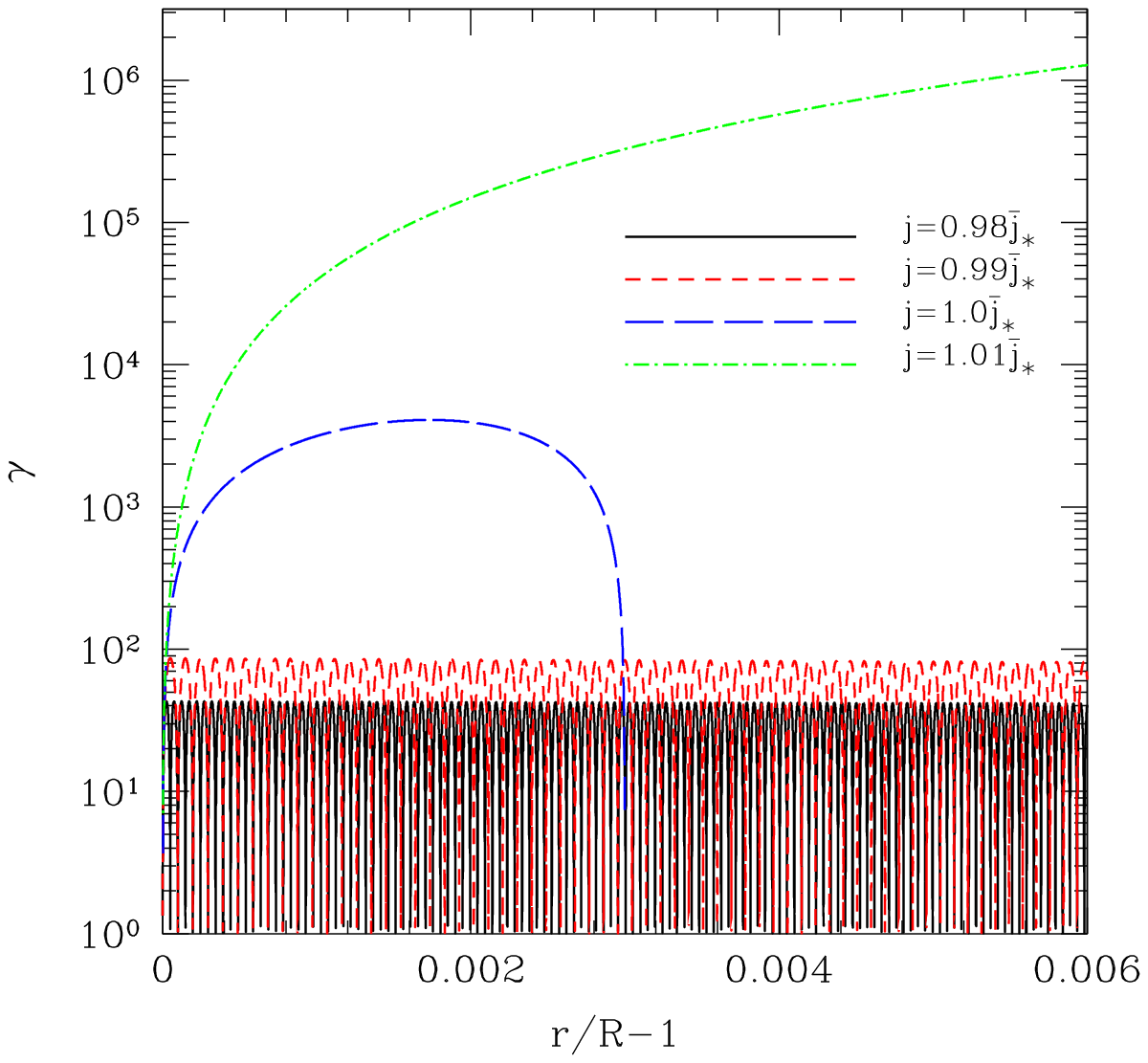}}
\vspace*{0.1cm}
\caption{Dependence of the Lorentz factor on the ratio $j/\bar{j}_\ast$
  for a neutron star with $M=1.4
  M_\odot$, $R=10 \ \mathrm{km}$, $P=0.1 s$, $\chi=30^\circ$,
  $B_0=1.0\times10^{12} \mathrm{G}$,
  $\theta_\ast=0^\circ$,  $\Theta_0=2^\circ$, and $\gamma_\ast=1.01$.
}
\label{fig0}
\end{figure}

\begin{figure}
\centering
{\includegraphics[angle=0,width=8.3cm,height=7.5cm]{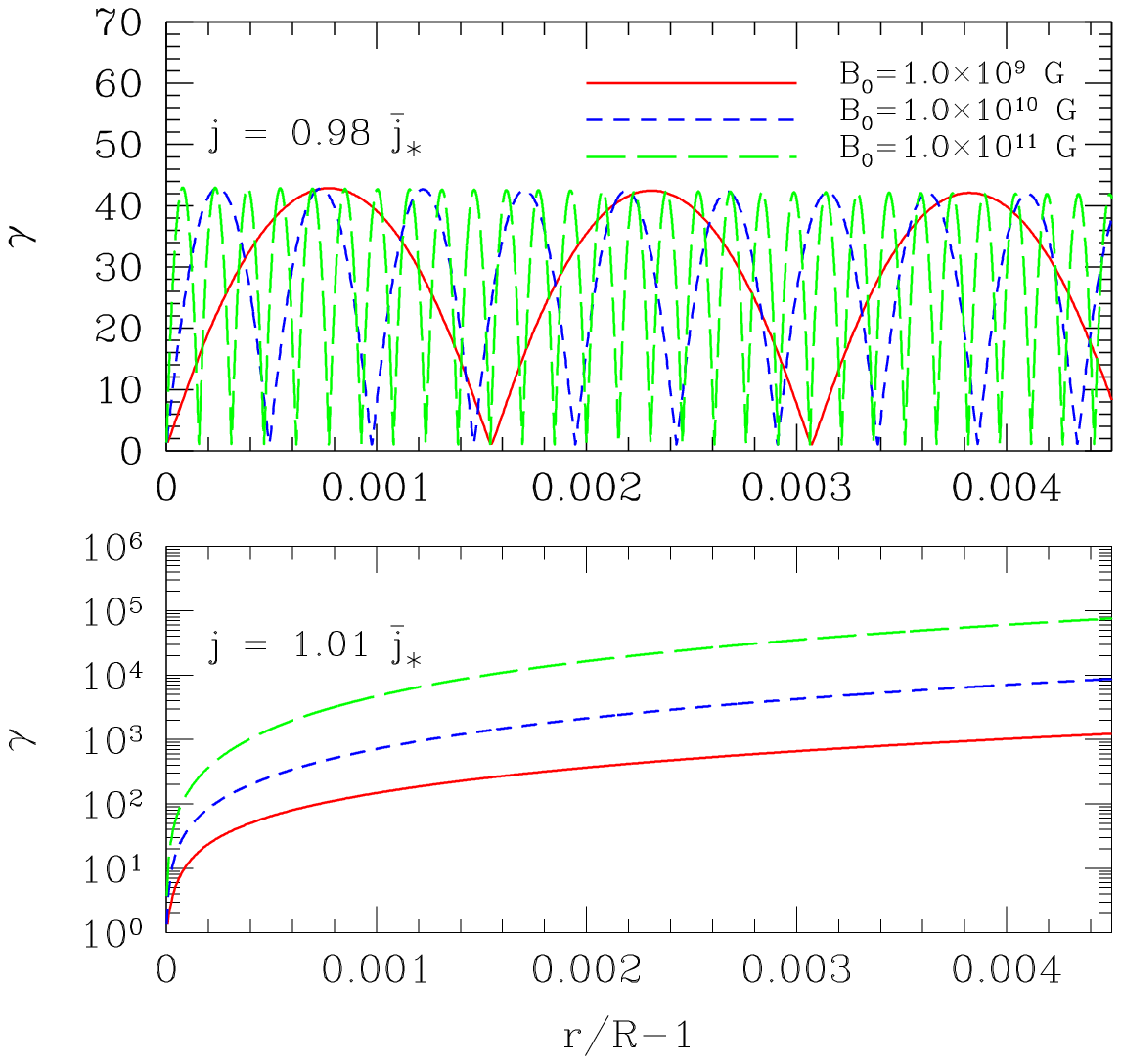}}
\vspace*{0.1cm}
\caption{Lorentz factor dependence on the intensity of
  the magnetic field for a neutron star with $M=1.4
  M_\odot$, $R=10 \ \mathrm{km}$, $P=0.1 s$, $\chi=30^\circ$,
  $\theta_\ast=0^\circ$, $\Theta_0=2^\circ$, $\gamma_\ast=1.01$.
  Top panel:
  $j=0.98\bar{j}_\ast$. Bottom panel: $j=1.01\bar{j}_\ast$.
}
\label{fig1}
\end{figure}

\begin{figure}
\centering
\vspace*{-3.5cm}
{\includegraphics[angle=0,width=8.3cm,height=7.5cm]{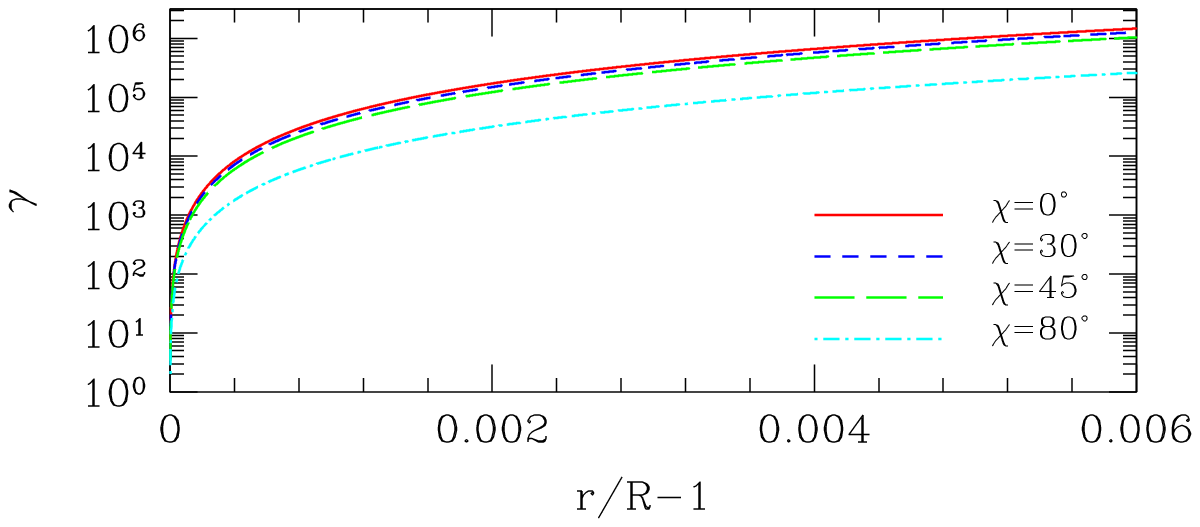}}
\caption{Lorentz factor dependence on the inclination
  angle $\chi$
  for a neutron star with $M=1.4
  M_\odot$, $R=10 \ \mathrm{km}$, and $P=0.1 s$, $j=1.01\bar{j}_\ast$,
  $\theta_\ast=0^\circ$, $\Theta_0=2^\circ$,
  $\gamma_\ast=1.01$, and
  $B_0=1.0\times10^{12} \mathrm{G}$. The Lorentz factor
  decreases for larger inclination angles.
}
\label{fig2}
\end{figure}

\begin{figure*}
\begin{center}
\includegraphics[angle=0,width=0.48\textwidth]{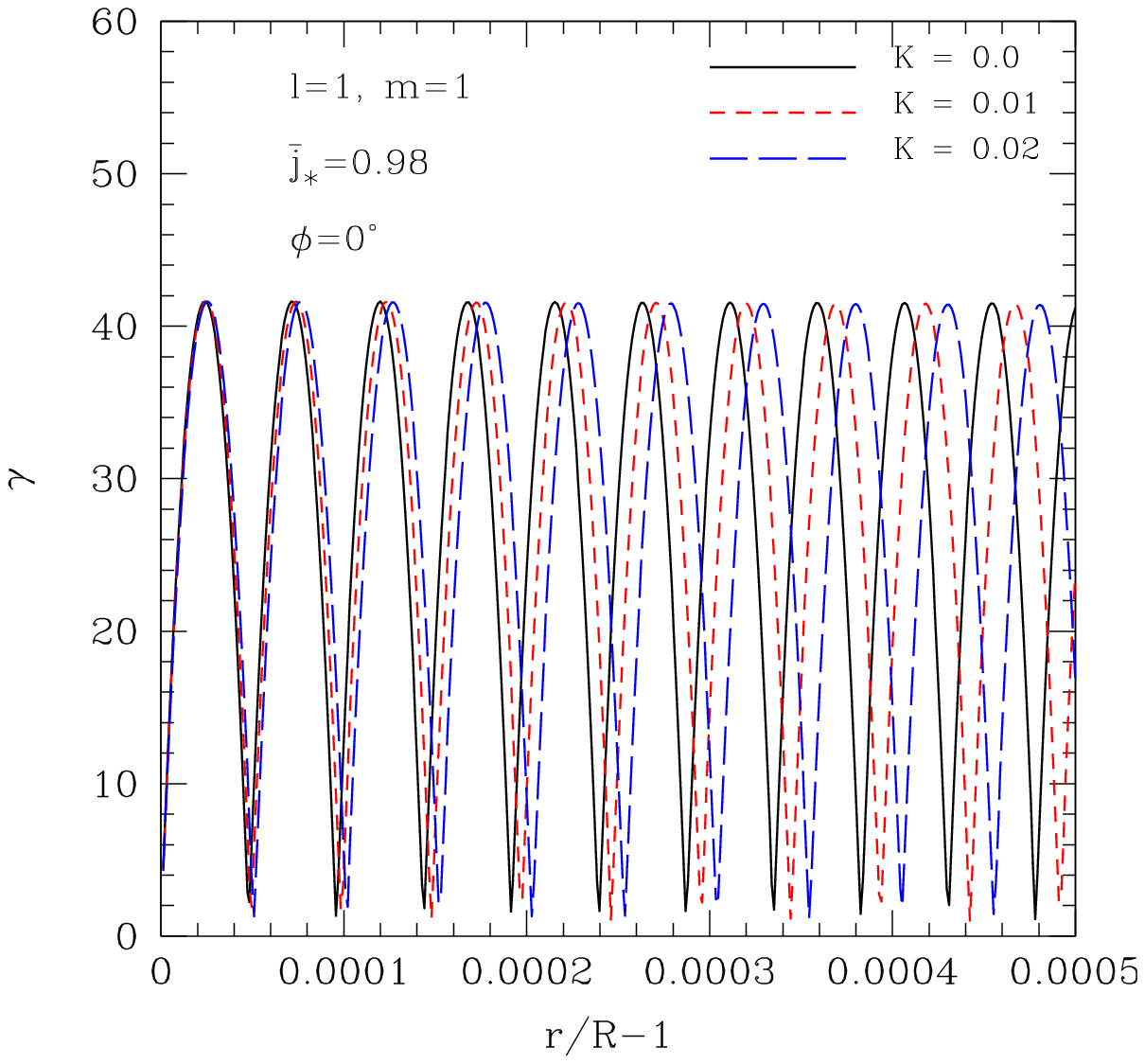}
\includegraphics[angle=0,width=0.48\textwidth]{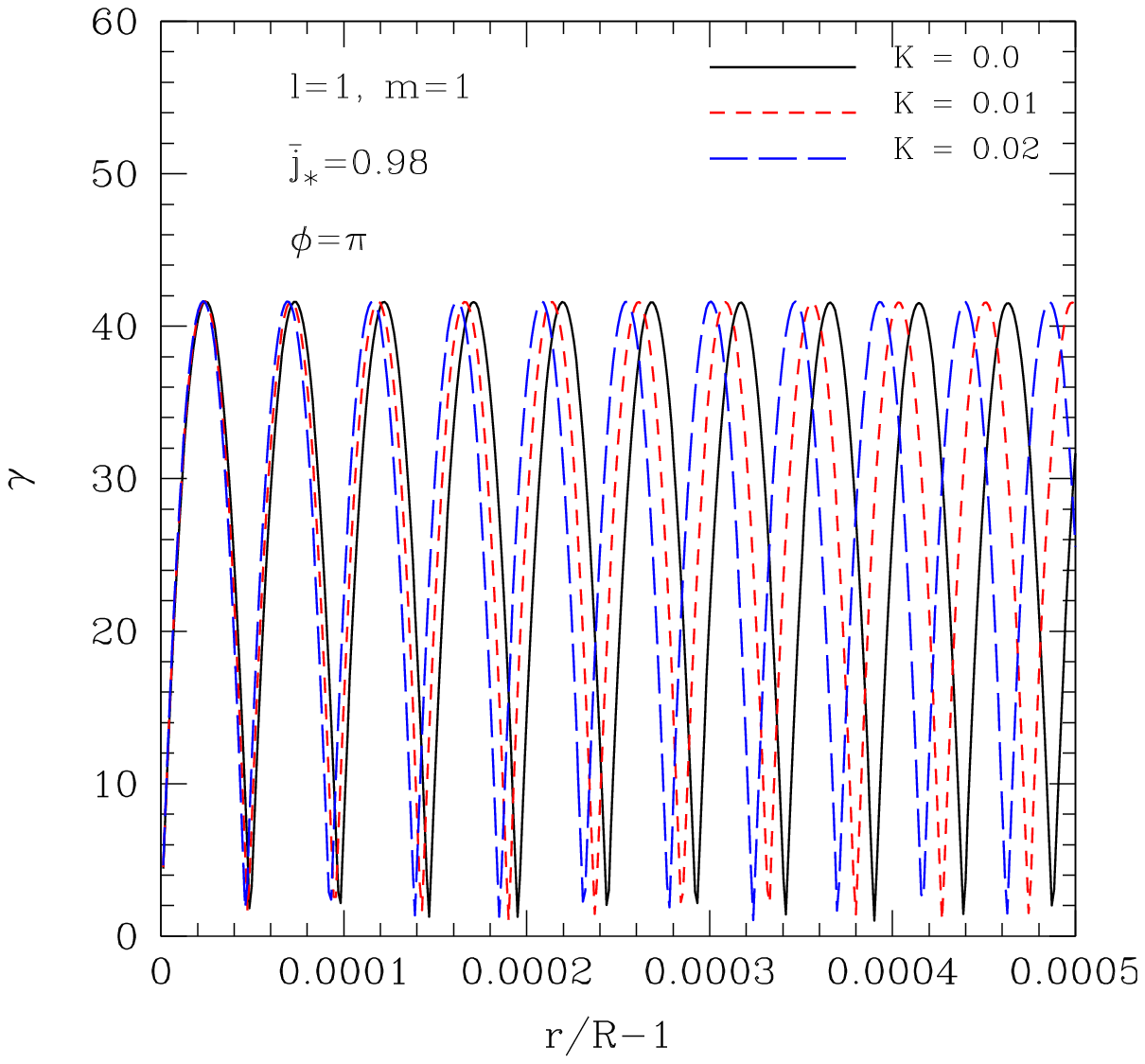}
\vspace*{0.1cm}
\includegraphics[angle=0,width=0.48\textwidth]{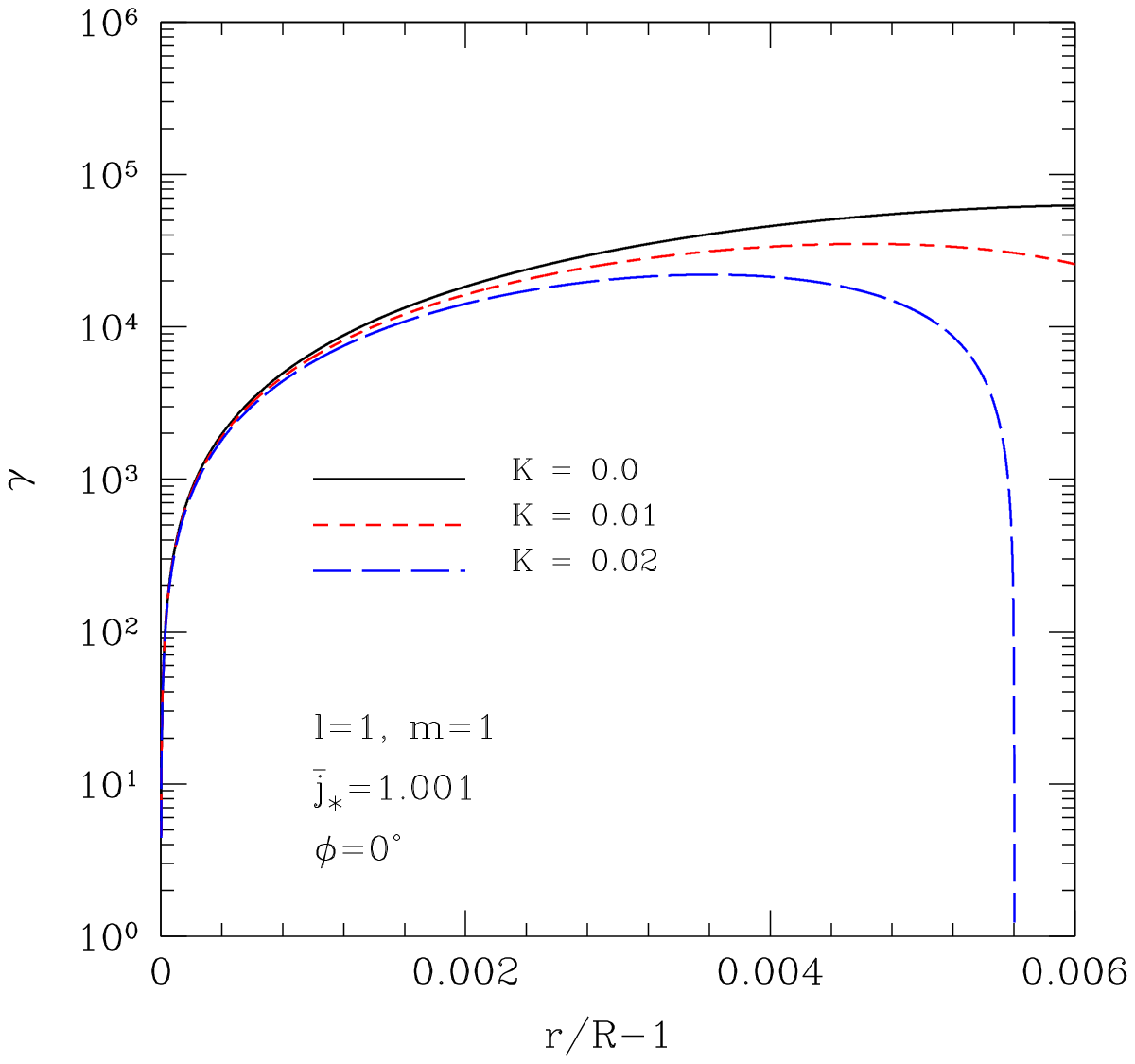}
\includegraphics[angle=0,width=0.48\textwidth]{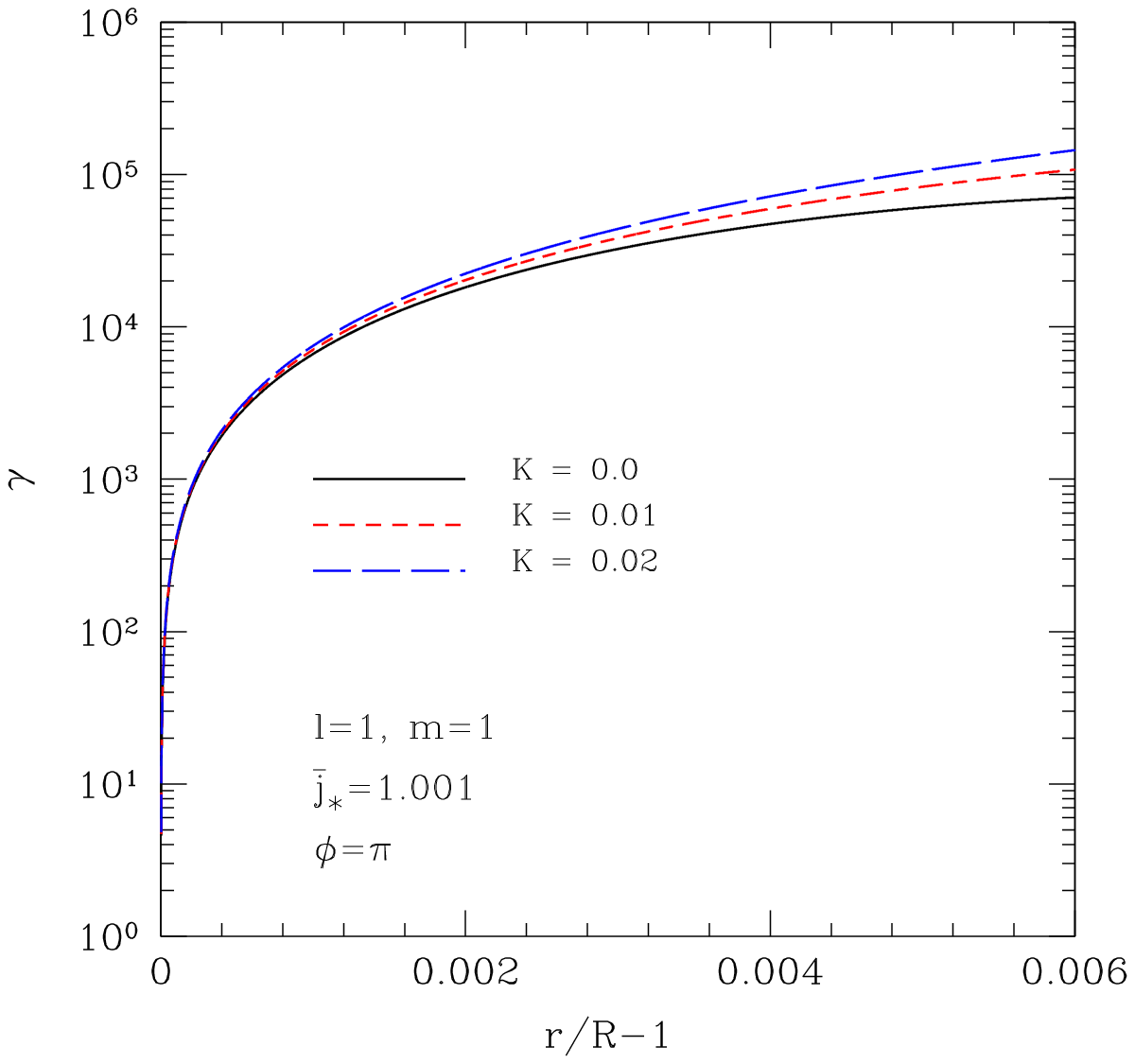}
\caption{Lorentz factor dependence on the normalized amplitude of
the stellar oscillations $K$ for the mode of oscillations
$(l,m)=(1,1)$ with
$\theta_\ast=2^\circ$,
$\Theta_0=3^\circ$,$\gamma_\ast=1.015$, and
  $B_0=1.0\times10^{12} \mathrm{G}$.
The two top panels correspond to the case
$j=0.98\bar{j}_\ast$, while the two bottom panels
correspond to the case $j=1.001\bar{j}_\ast$.
The left panels show the solution for $\phi=0$,
the right panels for $\phi=\pi$.}
\label{fig3}
\end{center}
\end{figure*}

\begin{figure*}
\begin{center}
\includegraphics[angle=0,width=0.48\textwidth]{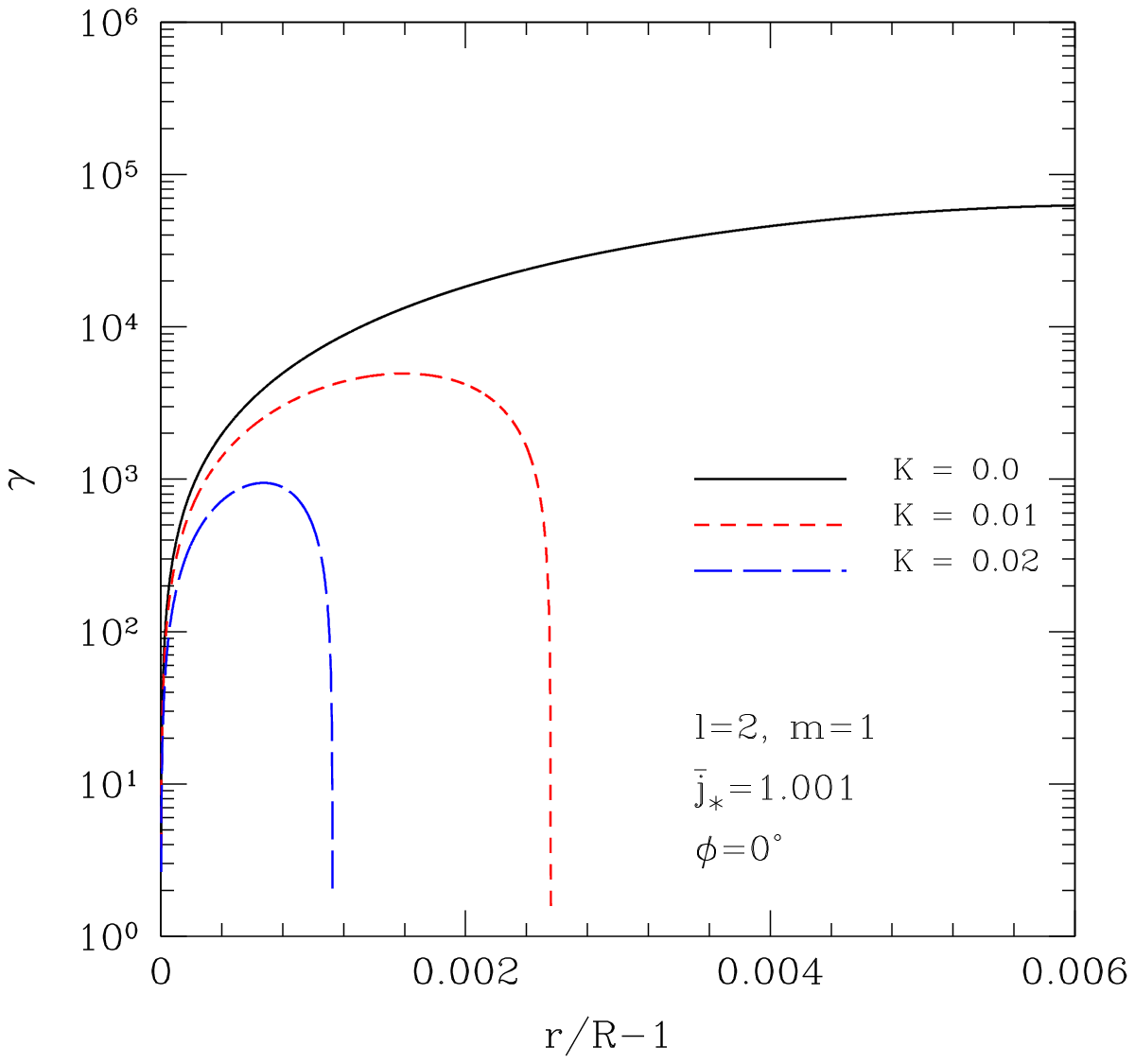}
\includegraphics[angle=0,width=0.48\textwidth]{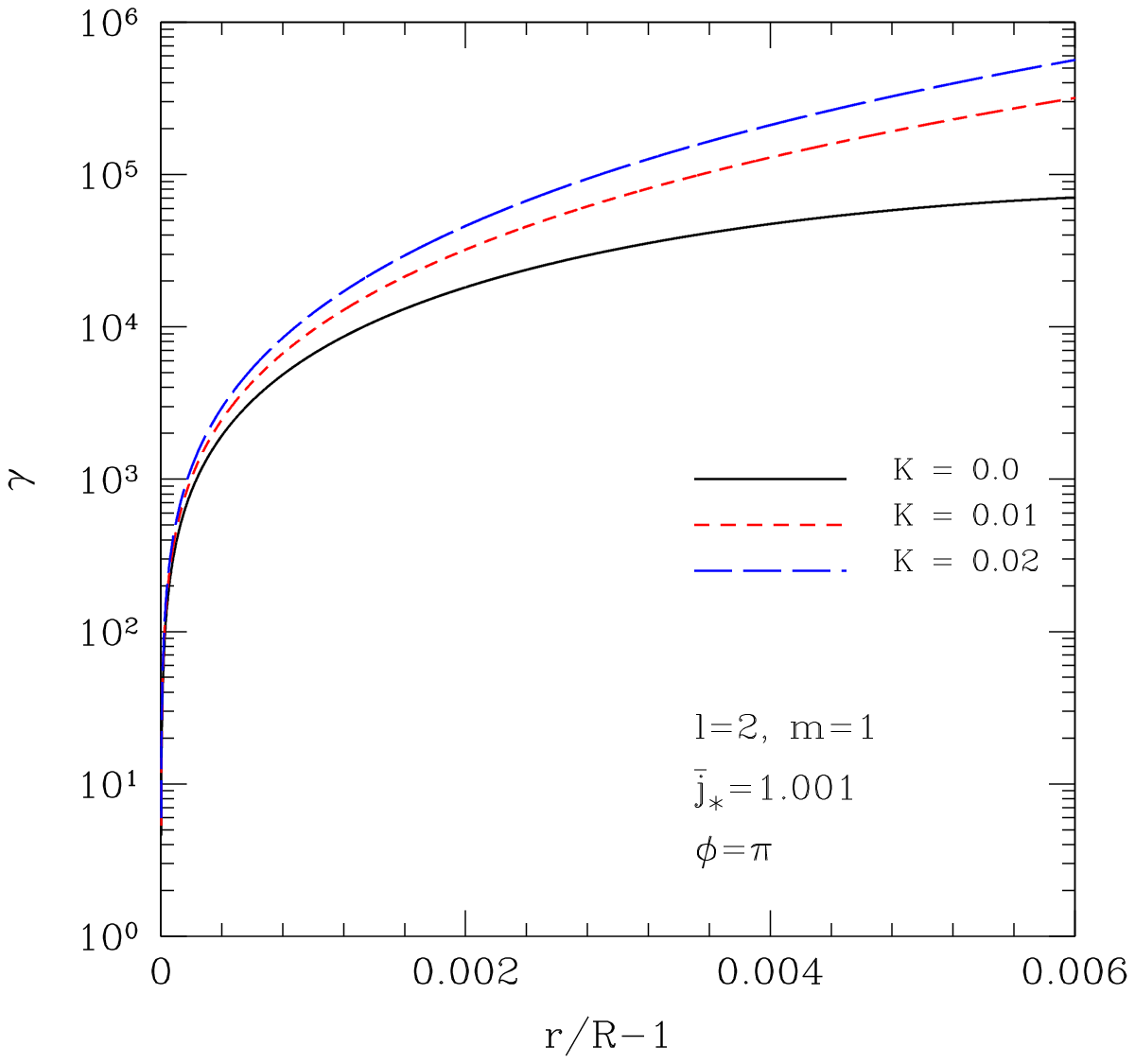}
\caption{
Lorentz factor dependence on the normalized amplitude of
the stellar oscillations $K$ for the mode of oscillations
$(l,m)=(2,1)$ with
$\theta_\ast=2^\circ$,
$\Theta_0=3^\circ$,$\gamma_\ast=1.015$, and
  $B_0=1.0\times10^{12} \mathrm{G}$.
The two panels
correspond to the case $j=1.001\bar{j}_\ast$.
The left panel shows the solution for $\phi=0$,
and the right panel for $\phi=\pi$.}
\label{fig4}
\end{center}
\end{figure*}

Fig.~\ref{fig0} reports the Lorentz factor as a function
of the normalized radial coordinate $r/R-1$ for
different values of the ratio $j/\bar{j}_\ast$.
Fig.~\ref{fig0} essentially confirms the
results of~\citet{Sakai2003}, which were originally
predicted by~\citet{Mestel1985},
 namely the existence of two different acceleration regimes.
When $j<\bar{j}$, small values of the velocity
$v$ can make the right-hand side of Eq.~(\ref{system1})
positive, thus producing an increasing potential and an
increasing Lorentz factor.
As soon as the velocity
reaches a critical value, given by the initial
ratio $j/\bar{j}$,
the right-hand side of Eq.~(\ref{system1})
becomes negative
and the potential, as well as the velocity, starts
to decrease. As a result, the case $j<\bar{j}$ is
responsible for the occurrence of an
oscillatory regime, with the oscillation
frequency (in the spatial domain) that is strongly dependent on the ratio
$j/\bar{j}_\ast$. As shown by \citet{Mestel1985} and \citet{Shibata1997}, this
situation corresponds to the case where the effective
charge density $\rho-\rho_{\rm GJ,rot}$ changes sign. When
$\rho-\rho_{\rm GJ,rot}>0$, the electron flow decelerates
to increase the charge density; in contrast, when
$\rho-\rho_{\rm GJ,rot}<0$, the flow accelerates  to
reduce the charge density.
A rather different behavior is
instead encountered when $j>\bar{j}$. In this case, 
the effective charge density is always negative,
the right hand side of Eq.~(\ref{system1})
is always positive, and the potential as well as
the  velocity can only grow.

\citet{Sakai2003}  erroneously
attributed the abrupt interruption of the oscillatory
pattern, which is manifested by some models
(see the bottom panel of their Fig. 1), to a purely general
relativistic effect.
In contrast and as we have verified,
this effect is purely numerical and caused by
$v^2$ possibly becoming negative during the
integration of the system of equations
Eqs.~(\ref{system1})-(\ref{system2}). 
For instance, this is also true
for the long-dashed blue curve in
Fig.~\ref{fig0}, where the Lorentz factor decays to unity at
$r/R-1\approx 0.003$ after having reached a value
of $\gamma_{max}\sim 4100$.
The pathology of the ODE
system that we have demonstrated
could not be solved even when resorting to stiff
solvers. As a result, the behavior of the Lorentz
factor after $v^2$ becomes negative cannot be trusted and
is not reported in our figures.
We also note that a value of $r/R-1=0.006$ corresponds to
a distance of $60 \ m$ above the star surface, which is
smaller than the radius of the polar cap $r_{\rm
  pc}\approx R ~ \Theta_0\approx 350 \ m$ as inferred
by the polar cap model of \citet{Ruderman1975}.

On the other hand, Fig.\ref{fig1}  illustrates the
dependence of the Lorentz factor of the accelerated particles on the
intensity of the magnetic field.
We note that
both the accelerating electric field $E_{\|}$
and the difference $\rho-\rho_{\rm GJ,rot}$
are directly proportional to the magnetic field~\citep[See
  Eq.~(56) by][]{Muslimov1992}.
Therefore,
if $j>\bar{j}_\ast$, corresponding to the case when the
effective charge density is always negative,
the Lorentz factor increases  with the
magnetic field, as shown in the bottom
panel of Fig.\ref{fig1}.
On the other hand,
if $j<\bar{j}_\ast$,
corresponding to the case when the
effective charge density changes sign, the charge density
jump $|\rho-\rho_{\rm GJ,rot}|$ increases with the magnetic
field, producing a pattern with a higher frequency of oscillation~\citep[See
  Eq.~(23) by][]{Shibata1997}. This effect is reported in
the top panel of Fig.\ref{fig1}.

\begin{figure*}
\includegraphics[angle=0,width=0.48\textwidth]{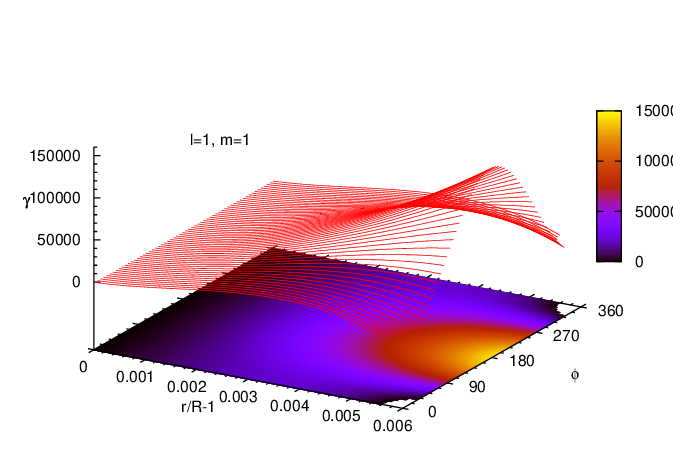}
\includegraphics[angle=0,width=0.48\textwidth]{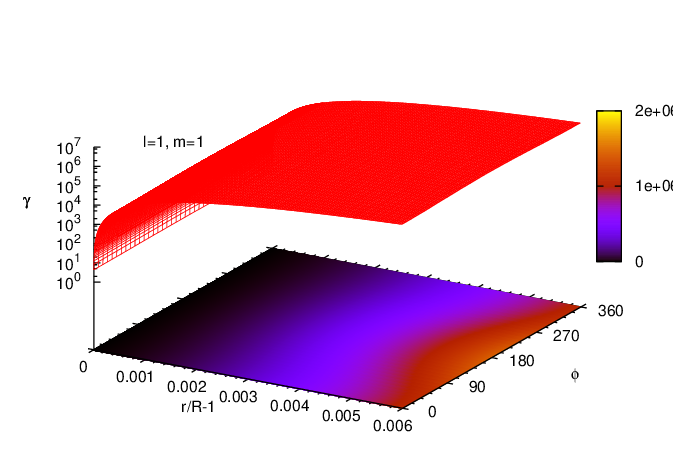}
\caption{
Lorentz factor as a function of radial distance and azimuthal angle $\phi$
for a model with stellar oscillations $K=0.02$,
$(l,m)=(1,1)$,
$\theta_\ast=2^\circ$,
$\Theta_0=3^\circ$,$\gamma_\ast=1.015$, and
  $B_0=1.0\times10^{12} \mathrm{G}$. Left panel:
$j=1.001\bar{j}_\ast$. Right panel: $j=1.01\bar{j}_\ast$.}
\label{fig5}
\end{figure*}

The dependence of the Lorentz factor on the inclination
angle $\chi$ is shown in Fig.\ref{fig2}.
When $j<\bar{j}_\ast$, namely when the oscillatory
pattern is produced, the Lorentz factor is 
insignificantly affected by variations in $\chi$ and the
results are not reported here.
In contrast, when $j>\bar{j}_\ast$,
the results are more interesting and
show that smaller Lorentz factors are obtained
for larger inclination angles $\chi$, although this
effect becomes significant only for very inclined
configurations.

In a second series of calculations,
we have considered the effects
of stellar oscillations
on the Lorentz-factor of the accelerated particles by
adopting the prescription given in Eq.~(\ref{barj}).
Figures~\ref{fig3} and ~\ref{fig4} show
the solutions for the modes of oscillations
$(l,m)=(1,1)$ and $(2,1)$, respectively.
As is apparent from the two top panels of
Fig.~\ref{fig3}, the
presence of stellar oscillations does not affect the amplitude of
the Lorentz factor of the particle for
$j=0.98\bar{j}_\ast$ and $(l,m)=(1,1)$.
On the other hand, the frequency
of the velocity oscillation is smaller for larger $K$ at
$\phi=0^\circ$ (left-top panel), while it is larger for larger $K$ at
$\phi=180^\circ$ (right-top-panel).
As long as the velocity of
oscillations for the considered modes is proportional to
$\cos\phi$, the effect will be the opposite when $\cos\phi$
changes sign.
The same qualitative behavior was also
encountered for the mode $(l,m)=(2,1)$,
and we conclude that it is typical of the case
$j/\bar{j}_\ast < 1$.

The case $j/\bar{j}_\ast\geq 1$ is reported in the two
bottom panels of Fig.~\ref{fig3} for the mode $(l,m)=(1,1)$
and in Fig.~\ref{fig4} for the mode $(l,m)=(2,1)$,
and illustrate that  star oscillations can significantly affect the
Lorentz factor.
We note that the influence of the
stellar oscillations on the acceleration of the particle
strongly depends on the azimuthal angle $\phi$.
In particular, owing to
the modulation produced by the term $\cos\phi$,
larger oscillation amplitudes $K$ become responsible 
  for a net
decelerating effect at angles $\phi\sim 0$, while they
produce a net accelerating effect at angles $\phi\sim \pi$.
Moreover, a strong dependence on the ratio
$j/\bar{j}_\ast$ is also encountered, as is clearly evident
from the two panels of Fig.~\ref{fig5}, which show
the modulation of the Lorentz factor with the angle
$\phi$ for the mode
$(l,m)=(1,1)$ in  a representative model with $K=0.02$.
The left panel refers to the case
$j/\bar{j}_\ast= 1.001$, while the right panel has $j/\bar{j}_\ast=
1.01$.
Interestingly,
\cite{Morozova2010} found that the oscillation modes with $m=1$ considerably
increase the electromagnetic energy losses from the polar cap region
of the neutron star, losses that can 
be several times larger than in the
case when no oscillations are present. 
A more efficient particle acceleration typically corresponds to
higher energy losses from the neutron star magnetosphere~\citep{Osmanov2009}.
As a result, we conclude that,
at least for the modes $(l,m)=(1,1)$ and $(l,m)=(2,1)$
that we have considered here, larger oscillation
amplitudes at the stellar surface may be responsible for both
stronger accelerations and higher luminosities, provided
that the current density in the magnetosphere is higher than
the Goldreich-Julian current density.

At least in the case of small inclination angles,
a noticeable influence of stellar oscillations on
the conditions for particle acceleration
can be intuitively understood as follows.
The acceleration of charged particles extracted from the surface of the
pulsar in the polar cap region is determined by the electric field,
which is generated by the combined effect  of strong 
magnetic fields and the motion of the star surface.
When stellar oscillations are added to the pure rigid
rotation, a given velocity distribution at the star
surface is produced.
The rotation velocity in the polar cap
region is proportional to the small polar angle $\theta$, is
exactly equal to zero on the axis of rotation, and increases with
distance away from the axis. 
In contrast, the velocity
of oscillations, which is given by Eq.~(\ref{vel}), in the case
$m=1$ does not depend on the small polar angle $\theta$
and is 
effectively constant across the polar cap region of the star. This
explains why for these modes even small oscillation velocities 
may have a noticeable effect on the accelerating component of the
electric field.

\section{CONCLUSIONS}
\label{Conclusions}

By  numerically solving the
relativistic electrodynamics equations in the stationary
regime of a space-charge-limited-flow accelerator,
we have computed the Lorentz factor of
electrons accelerated in the polar cap region of a
rotating and oscillating pulsar magnetosphere.
Our results confirm some of the fundamental findings of
\citet{Sakai2003},
namely the existence of two different
regimes: an oscillatory one, which is produced for
sub-GJ current density configurations,
($j/\bar{j}_\ast<1$) and does not produce an
efficient acceleration, and a truly accelerating one,
which is produced for super-GJ current density
configurations ($j/\bar{j}_\ast>1$).
We have also clarified the numerical origin of a stopping
effect on the velocity of the electron,
which has nothing to do with general relativistic effects.
Stellar oscillations, on the other hand,  can affect
both the absolute value of the Lorentz factor gained
by the electron and the frequency of the oscillatory
patterns, when the latter is present.
Owing to the modulation produced by the term $\cos\phi$ in the
equations, where $\phi$ is the azimuthal angle,
the way in which each  mode of oscillation affects the dynamics
is different at different
azimuthal positions on the polar
cap surface, and, in particular, it follows the
periodicity of $\cos\phi$.
These results are overall consistent with those
obtained by \citet{Morozova2010}, who showed
that oscillations are responsible for an extra term
in the total energy losses from the system, because the electromagnetic
energy losses are determined by the integrated absolute value of the
current, flowing in the magnetosphere.
At least for the modes $(l,m)=(1,1)$ and $(l,m)=(2,1)$,
larger oscillation
amplitudes at the star surface may produce both higher
accelerations and higher luminosities, provided
that the magnetosphere has a super-GJ current density
configuration.
However, these results also suggest
that pulsar
oscillations may become responsible for
a significant asymmetry in the pulse
profile, which will depend on the orientation of the oscillations with
respect to the pulsar magnetic field.

Finally, we stress that some recent investigations have tried
to connect the models of stellar oscillations with the observational
data available for pulsars \citep{Rosen2008,Rosen2011a,Rosen2011b}.
The scenario that is emerging from these studies is that the presence of stellar
oscillations creates different kinds of "noise" in the
clock-like picture of pulses, i.e. changes in the pulse shape,
changes in the spin-down rate, and the switching between different
regimes of pulsar emission. 
\cite{vanLeeuwen2012}, in particular, 
revisited the issue of drifting subpulses in the pulsar
profiles, reaching the important conclusion 
that the rate of subpulse drift is proportional to the
latitudinal derivative of the scalar potential across the pulsar
polar cap and not to the absolute value of the scalar potential, as
had been generally assumed.
This provides a wide range of opportunities
for  comparing the observational data on pulsar profiles
with the results presented here and in our previous
investigations of oscillating magnetopsheres.
Stellar oscillations, indeed, affect the scalar potential in the polar cap
region and are expected to have a significant effect on the character
of the subpulse drift. We plan to devote our future research to a
closer analysis of the observational aspects of these ideas.

\begin{acknowledgements}
V.Morozova would like to acknowledge the support of the German
Academic Exchange Service DAAD for supporting her stay at ZARM.
\end{acknowledgements}

\appendix
\section[]{Extended geometrized system of units}
\label{appendixA}

We recall that the definition of geometric units of time and lengths
is obtained by setting the speed of light $c$ and the gravitational
constant $G$ to pure numbers. This implies that seconds and grams of
the $\rm {cgs}$ system can be written as
\begin{eqnarray}
\label{second}
1 {\rm s}&=&2.997924 {\times} 10^{10}\,\left(\frac{1}{c}\right)\, \rm{cm} \,  \\
\label{gram}
1 \rm{g}&=&7.424157{\times} 10^{-29}\,\left(\frac{c^2}{G}\right)\, \rm{cm} \,.
\end{eqnarray}
Within this general setup, a convenient unit of space is still
required. The $\rm{cm}$ is of course a bad choice in a
gravitational framework and the
gravitational radius $r_g=G M/c^2$ is instead chosen. In order for
this new unit to be more convenient than the centimeter, the
mass $M$ of the system has to be sufficiently high.  From the
physical value of the solar mass and Eq.~(\ref{gram}), we find the
relations between the cgs units and the new unit of length $r_g$
\bea
\label{centimeter_2}_
1\rm{cm}&=&6.772289 {\times} 10^{-6}\,
\left(\frac{M_{\odot}}{M}\right)\, r_g\,, \\
\label{second_2}
1\rm{s}&=&2.030281 {\times} 10^{5} \,
\left(\frac{1}{c}\right) \left(\frac{M_{\odot}}{M}\right)\,r_g\,, \\
\label{gram_2}
1\rm{g}&=&5.027854{\times} 10^{-34} \,
\left(\frac{c^2}{G}\right) \left(\frac{M_{\odot}}{M}\right)\,  r_g \,.
\eea
In the traditional geometrized system $c$ and
$G$ are set equal to unity.  The equations
Eqs.~(\ref{centimeter_2})-(\ref{gram_2}) allow us to
perform the
geometrization of any dynamical quantities. Quantities
involving the charge, on the other hand, are naturally
geometrized within the Gauss system of units, which we
adopt here in addition to the choice $c=G=1$.
In Gauss units, the unit of
measure of the magnetic induction, the Gauss G, is given
by
\be
1G=\rm{g}^{1/2} \rm{cm}^{-1/2}\rm{s}^{-1} \ ,
\ee
which, by virtue of (\ref{centimeter_2})-(\ref{gram_2}),
provides
\be
B_{\rm{geo}}=4.2439\times
10^{-20}\left(\frac{M}{M_{\odot}}\right)B_{\rm{G}} \ ,
\ee
where $B_{\rm{G}}$ and $B_{\rm{geo}}$
are the pure numbers expressing
the intensity of the magnetic field
in Gauss units and the geometrized
system, respectively. We note that in
these units the charge-to-mass ratio is a pure
number, and, in the case of electrons, is given by
\be
\frac{e}{m_e}=2.0425\times10^{21} \ .
\ee

\bibliographystyle{aa}


\end{document}